\RequirePackage{fix-cm}
\documentclass[smallextended]{svjour3}       
\smartqed  
\usepackage{graphics,epstopdf}
\usepackage{graphicx}
\usepackage[colorlinks=true,citecolor=blue, urlcolor=blue]{hyperref}
\usepackage{float}

\begin{document}
\title{Uncertainty principle guarantees genuine source of intrinsic randomness}
\author{Trina Chakraborty         \and
        Manik Banik \and
        Pinaki Patra 
}
\institute{Trina Chakraborty \at
              Physics and Applied Mathematics Unit, Indian Statistical Institute, 203 B.T. Road, Kolkata-700108, India\\
              \email{trinachakraborty.27@gmail.com}
              \and
            Manik Banik \at
              Physics and Applied Mathematics Unit, Indian Statistical Institute, 203 B.T. Road, Kolkata-700108, India\\
              \email{manik11ju@gmail.com}           
           \and
          Pinaki Patra\at
              Department of Physics, University of Kalyani, India-741235\\
                            \email{monk.ju@gmail.com}           
}

\date{Received: date / Accepted: date}
\maketitle
\begin{abstract}
The Born's rule introduces intrinsic randomness to the outcomes of a measurement performed on a quantum mechanical system. But, if the system is prepared in the eigenstate of an observable then the measurement outcome of that observable is completely predictable and hence there is no intrinsic randomness. On the other hand, if two incompatible observables are measured (either sequentially on a particle or simultaneously on two identical copies of the particle) then uncertainty principle guarantees intrinsic randomness in the subsequent outcomes independent of the preparation state of the system. In this article we show that this is true not only in quantum mechanics but for any no-signaling probabilistic theory. Also the minimum amount of intrinsic randomness that can be guaranteed for arbitrarily prepared state of the system is quantified by the amount of (un)certainty.
\end{abstract}

\section{Introduction}
Heisenberg's Uncertainty Principle \cite{heisenberg} is one of the primitive constitutional concepts of Quantum Physics. It makes a fundamental difference between quantum theory of physical world and it's classical counterpart and drastically modifies our classical conceptual framework. The uncertainty principle states that there are incompatible measurements, such as position and momentum, for which there is a trade-off relationship in the degrees of sharpness of the preparation or measurement of their values, such that a simultaneous or sequential determination of the values requires a nonzero amount of unsharpness \cite{busch1,busch2}. This principle actually states a fundamental property of quantum systems, and is not a statement about the observational limitation of current technology. 

Along with uncertainty principle, Born's rule is another important key aspect in quantum mechanics, first stated by Max Born in the context of scattering theory \cite{born}. This rule provides a link between the mathematical formalism of quantum theory and experiment and almost single-handedly responsible for practically all predictions of quantum physics. 

In the history of science, Born's rule is often seen as a turning point where intrinsic randomness entered into fundamental physics. Note that if the system is prepared in one of the eigenstates of a given observable then the outcome of the given observable is fully deterministic. Thus given an observable acting on a quantum system one can not associate  intrinsic randomness to the outcomes independent of preparation state of the system. In other word, Born's rule cannot guarantee intrinsic randomness to the outcomes of a physical process for arbitrarily prepared state of a quantum system. In contrast to this, if two incompatible observables are measured either sequentially on a particle or simultaneously on two identical copies of the particle then according to the uncertainty principle intrinsic randomness is associated with the subsequent outcomes independent of the prepared state of the quantum system. 

Here we show that existence of such preparation state independent random process is guaranteed by uncertainty principle not only quantum theory but in any general no signaling theory (GNST). Considering a particular form of uncertainty principle, namely fine-grained uncertainty relation recently introduced in \cite{oppenheim}, we show that the minimum amount of intrinsic randomness that can be guaranteed for arbitrarily prepared state of the given quantum system is determined by the amount of (un)certainty present in quantum mechanics. This quantitative link also holds in all other GNSTs.

\section{Fine-grained uncertainty relation}
Measurements allow us to gain information about the state of a physical system. In quantum mechanics uncertainty principle imposes some limitation on what we can hope to learn about the state of the system. The general form of the Heisenberg's uncertainty relation for two observables $A$ and $B$, introduced by Robertson \cite{robertson}, looks
\begin{equation}
\Delta A \Delta B \geq \frac{1}{2}|\langle\psi|[A,B]|\psi\rangle|
\end{equation}
where $\Delta X = \sqrt{\langle\psi|X^2|\psi\rangle-\langle\psi|X|\psi\rangle^2}$ represents the standard deviation which is a measure of uncertainty of the corresponding observable $X$, for $X\in \{A,B\}$. For many situations the standard deviation is not a natural way of quantifying uncertainty \cite{carruthers,louisell}. A modern approach to overcome these issues is to consider entropic measure for quantifying uncertainty \cite{bialynicki}. We use the notation $p_\tau(o^{(m)}|m)$ to denote the probability of obtaining outcome $o^{(m)}$ when a measurement $m$, chosen from a set of measurements $\mathcal{M}$, is  performed on a system in state $\tau$. In quantum theory, the state of a system is described by a density operator acting on a Hilbert space, while for a general theory, one can consider $\tau$ as an abstract representation of a state. The Shannon entropy of the distribution over measurement outcomes of measurement $m$ on a system in state $\tau$ is given by
\begin{equation}
H_\tau(m)=-\sum_{o^{(m)}}p_\tau(o^{(m)}|m) \log_2~p_\tau(o^{(m)}|m)
\end{equation}
A general entropic uncertainty relation is of the form
\begin{equation}
\sum_mp(m)H_\tau(m)\geq c_{\mathcal{M},\mathcal{D}}
\end{equation}
where $p(m)$ is any probability distribution over the set of measurements $\mathcal{M}$, and $c_{\mathcal{M},\mathcal{D}}$ is some positive constant determined by $\mathcal{M}$ and the distribution $\mathcal{D}=\{p(m)\}_m$. Please note that, the lower bound $c_{\mathcal{M},\mathcal{D}}$ is independent of the state $\tau$. Entropic uncertainty relations for two observables was first introduced by Deutsch \cite{deutsch}, then a improved version was conjectured \cite{kraus} and then proved \cite{maassen} (see also \cite{wehner} for a recent survey about entropic uncertainty relations and references therein). 

In \cite{oppenheim}, the authors pointed out that entropic functions are a coarse way of measuring the uncertainty of a set of measurements as they do not distinguish the uncertainty inherent in obtaining any combination of outcomes $o(m)$ for different measurements $m$. Thus they introduce fine-grained uncertainty relations consisting of a series of inequalities, one for each combination of possible outcomes, which can be written as a string $\vec o=(o^{(1)},...,o^{(n)})\in\mathcal{B}^{\times n}$ with $n=|\mathcal{M}|$. That is, for each $\vec o$, a set of measurements $\mathcal{M}$ , and distribution $\mathcal{D} = \{p(m)\}_m$
\begin{equation}
P^{cert}(\tau;\vec o) := \sum_{m\in\mathcal{M}}p(m) p_\tau(o^{(m)}|m)\leq\zeta_{\vec o}(\mathcal{M},\mathcal{D})
\end{equation}
For a fixed set of measurements, the set of inequalities
\begin{equation}
\mathcal{U}=\{ \sum_{m\in\mathcal{M}} p_\tau(o^{(m)}|m)\leq\zeta_{\vec o}~|~\forall \vec o\in \mathcal{B}^{\times n}\}
\end{equation}
thus forms a fine-grained uncertainty relation. These relations dictates that one cannot obtain a measurement outcome with certainty for all measurements simultaneously whenever $\zeta_{\vec o}<1$. In this fine-grained version of the uncertainty relation, the amount of (un)certainty in a particular theory is characterized by the values of
\begin{equation}
\zeta_{\vec o}=\max_{\tau} \sum_{m\in\mathcal{M}}p(m) p_\tau(o^{(m)}|m)
\end{equation}
where the maximization is taken over all states allowed for a particular system in the concerned theory.
\section{ Genuine randomness source in GNST}
Studying physical theory in general probabilistic framework has motivated recently by the work of L. Hardy \cite{hardy}, and many interesting research works have been done in this field \cite{barrett,barnum1,barnum2,barnum3,barnum4,barnum5,chiribella,masanes,pfister}. Let $\tau$ describe the state of a system in general  theory (GNST) which belongs in a convex state space $\Gamma$. Convexity of the state space $\Gamma$ implies that any probabilistic mixture of two states is again a possible state of the system. Given any state $\tau\in\Gamma$, a GNST assigns a probability measure $p_\tau(a|A)$ for obtaining outcomes $a\in\{a_1,...,a_n\}$ when measurement $A$ is performed on the system. If, for a given $\tau$, $p_\tau(a|A)$ is different from $1$ for all $a$, then we can say that the measurement process $A$ induce an intrinsic randomness to its outcomes in the concerned GNST, whenever the state of the system is described by $\tau$. We can quantify the randomness of the outcome $a$ resulting from the measurement of observable $A$ on a state $\tau$ through the guessing probability \cite{pironio,acin}
\begin{equation}
G(\tau,A)=\max_a~ p_\tau(a|A)
\end{equation}
The guessing probability can be expressed in bits and is then known as the min-entropy \cite{koenig}
\begin{equation}
H_\infty(\tau,A)=-\log_2G(\tau,A)
\end{equation}
Though measurement $A$ induces intrinsic randomness to it's outcomes when the system's state is $\tau$, it does not give security of intrinsic randomness for arbitrarily prepared state of the system. Thus the measurement process $A$ may not be a \emph{genuine source of intrinsic randomness} (defined following) in the concerned GNST.

\textbf{Definition :} \emph{In any GNST, a physical process will be called a genuine source of intrinsic randomness if it guarantees nonzero amount of intrinsic randomness for every possible system's states allowed in the concerned GNST.}

It may happen that the outcomes of the measurement $A$ are random when the system is in the state $\tau_1$, whereas the outcomes are deterministic when the state of the system is $\tau_2$. As for example in quantum spin-$\frac{1}{2}$ system the outcomes of spin measurement along z-direction, according to Born's rule, is fully random if the system is in one of the the eigenstates of $\sigma_x$ or $\sigma_y$; but the same Born's rule assigns deterministic outcomes for $\sigma_z$ measurement if the system is in one of it's eigenstates. Actually for any measurement process in quantum mechanics  outcomes are deterministic when the system is in one of it's eigenstates. Thus according to our definition, quantum mechanical measurement processes are not \emph{genuine source of intrinsic randomness}. Let there arises a situation that one, say Alice, has to produce some (private) randomness by performing a (publicly) known measurement on many copies of identically prepared quantum mechanical system, but an un-trusted party, say Bob, makes supply (with no quantum memory \cite{berta,prevedel}) of the quantum system to her. In this situation the postulated Born's rule, alone, cannot guarantee the desired randomness as Bob might prepare the system in one of the eigenstates of the measurement observable. Interestingly in the following we show that the uncertainty principle, independent of any further assumption, can guarantee such randomness not only in quantum mechanics but in all GNST.

\section{ Uncertainty guarantees genuine source of randomness}
Consider two two-outcomes incompatible measurements, say $m_1$ and $m_2$, in a GNST with system's state space $\Gamma$. Taking $\mathcal{D}$ uniform, let the fine-grained uncertainty relation \cite{oppenheim} confirms $\zeta_{\vec{o}, m_1, m_2}$ amount of (un)certainty in the concerned GNST $(\vec{o}\equiv(o^{(m_1)}, o^{(m_2)})$, where $\zeta_{\vec{o}, m_1, m_2}$ is given by
\begin{equation}
\zeta_{\vec{o}, m_1, m_2}=\max_{\tau\in\Gamma}~[~\frac{1}{2} p_\tau(o^{(m_1)}|m_1)+\frac{1}{2} p_\tau(o^{(m_2)}|m_2)~]
\end{equation}
For notational simplicity from now on we denote outcomes of both the measurements $m_1$ and $m_2$ by $0$ and $1$, i.e. $o^{(m_1)},o^{(m_2)}\in\{0,1\}$. With this assumed amount of (un)certainty in our hand in the concerned GNST, we state our main result in the following theorem.

\textbf{Theorem :} \emph{If in any GNST the amount of (un)certainty amounts to $\zeta$, then in the concerned GNST there exists a genuine random process which guarantees at least $-2\log_2\zeta$ bits of intrinsic randomness.}

\textbf{Proof :} Given many copies of identically prepared system, in any allowed state $\tau\in\Gamma$, we perform measurement $m_1$ on the $1^{st}$ copy of the system producing outcomes $0$ and $1$ with probabilities $p_\tau(0|m_1)$ and $p_\tau(1|m_1)$ respectively. Measurement $m_2$ on the $2^{nd}$ copy produces outcomes $0$ and $1$ with probabilities $p_\tau(0|m_2)$ and $p_\tau(1|m_2)$ respectively. As these two measurement processes are independent, it can be considered as a four outcomes process with outcomes denoted by $(j,k)$ with occurrence probabilities $p_\tau(j|m_1)p_\tau(k|m_2)$ respectively, where $j,k\in\{0,1\}$. This process is repetitive. Among these four probabilities min-entropy of the highest one quantifies (eqn.(8)) the random bits associated with the process. Without loss of generality consider that maximum (un)certainty is achieved for the pair (0,0), i.e.
\begin{equation}
\frac{1}{2} p_\tau(0|m_1)+\frac{1}{2} p_\tau(0|m_2)\leq\zeta~~~~\forall~\tau\in\Gamma
\end{equation}
$p_\tau(0|m_1)$ and $p_\tau(0|m_2)$ are real number lying in the interval $[0,1]$, we thus have
\begin{eqnarray*}
(\sqrt{p_\tau(0|m_1)}-\sqrt{p_\tau(0|m_2)})^2\geq0\\
\Rightarrow~~2\sqrt{p_\tau(0|m_1)p_\tau(0|m_2)}\leq p_\tau(0|m_1)+p_\tau(0|m_2)\\
\Rightarrow~~p_\tau(0|m_1)p_\tau(0|m_2)\leq \frac{1}{4}\{p_\tau(0|m_1)+p_\tau(0|m_2)\}^2\\
\Rightarrow~~p_\tau(0|m_1)p_\tau(0|m_2)\leq \zeta^2~~~~\forall~\tau\in\Gamma
\end{eqnarray*}
Therefore the process described above associates $-2\log_2\zeta~(=-\log_2\zeta^2)$ bits of  intrinsic randomness to the outcome string. Moreover this amount of intrinsic randomness is associated for any preparation state of the system. Thus uncertainty principle guarantees existence of a genuine random process--hence the theorem follows.

\section{ Genuine source of randomness in different GNSTs}
Given a GNST, with system's state space $\Gamma$, we can construct hidden variable theory (HVT) \cite{ariano} (see also \cite{brandenburger} for very interesting discussion about HVT). In this HVT, state of the system is described by $\tau$ combined with another parameter $\lambda\in\Lambda$. Let's denote the state space of the system in this HVT as $\Gamma\times\Lambda$.
Given the knowledge of the system specified by the pair $(\tau,\lambda)\in\Gamma\times\Lambda$, the HVT assigns a probability rule $p_{(\tau,\lambda)}(o^{(m)}|m)$ for obtaining outcomes $o^{(m)}$ when measurements $m\in\mathcal{M}$ is performed on the system. Let the amount of uncertainty in our concerning GNST and the corresponding HVT is quantified by $\zeta_{\Gamma}$ and $\zeta_{\Gamma\times\Lambda}$ respectively. The minimum amount of intrinsic randomness that can be guaranteed for arbitrarily prepared state in the concerned GNST and the corresponding HVT is quantified by our theorem accordingly. In some cases, in principle, it is possible to construct realistic HVT where there is no uncertainty and therefore intrinsic randomness vanishes in this realistic HVT. In the following we discuss few important theories.

\emph{Classical Physics} : In classical physics there is no uncertainty relation between any pair of observables i.e. for any pair of observables in classical mechanics we always have $\zeta_{cl}=1$. Therefore in classical mechanics we cannot have \emph{genuine source of intrinsic randomness}.

\emph{Quantum mechanics} : As discussed in \cite{oppenheim} the amount of (un)certainty in quantum mechanics amounts to $\zeta_Q=\frac{1}{2}+\frac{1}{2\sqrt{2}}$. According to our derived formula there exists a genuine random process which certifies at least $-2\log_2(\frac{1}{2}+\frac{1}{2\sqrt{2}})\cong0.457$ bits of  intrinsic randomness. Performing $\sigma_z$ and $\sigma_x$ measurements on the $1^{st}$ and $2^{nd}$ copy of the system respectively this minimum bound of intrinsic randomness can be achieved when the state of the system is one of the eigenstats of $(\sigma_z\pm\sigma_x)/\sqrt{2}$. In quantum mechanics the same process may allow at most $2$ bits of intrinsic randomness, which is achieved when the system's is prepared in one of the eigenstates of $\sigma_y$. For other choices of quantum states intrinsic randomness lies between these two extreme values.

\emph{Bell-Mermin model for 2-states quantum system} : An ontological model for a two dimensional Hilbert space has originally introduced by Bell \cite{bell} and then Mermin presented it in a more intuitive form \cite{mermin} (see \cite{spekkens} for a quick view of this model). The model employs an ontic state space $\Lambda$ that is a Cartesian product of a pair of state spaces, $\Lambda=\Lambda'\times\Lambda''$. Each of $\Lambda'$ and $\Lambda''$ is isomorphic to the unit sphere. A system prepared according to quantum state $\psi$ is assumed to be described by a product distribution $p(\vec{\lambda'},\vec{\lambda''}|\psi)=p(\vec{\lambda'}|\psi)p(\vec{\lambda''}|\psi)$ on $\Lambda'\times\Lambda''$, where $p(\vec{\lambda'}|\psi)$ is a Dirac-delta function centered at $\psi$ and $p(\vec{\lambda''}|\psi)$ is a uniform distribution independent of $\psi$. When projective measurement associated with the basis $\{\phi,\phi^{\bot}\}$ is performed Bell-Mermin model associated this measurement with the indicator function
\begin{equation}
p(\phi|\vec{\lambda'},\vec{\lambda''})=\Theta(\vec{\phi}.(\vec{\lambda'}+\vec{\lambda''}))\nonumber\\
\end{equation}
where $\Theta$ is the Heaviside step function defined by
\begin{eqnarray}
\Theta(x)=1~if~x>0\nonumber\\
=0~if~x\leq0\nonumber
\end{eqnarray}
This Heaviside step function clearly indicates that Bell-Mermin model is a realistic HVT of $2$-states quantum system, and in this theory we have no uncertainty as well as no \emph{genuine source of intrinsic randomness}.

\emph{Box world (PR-box)} : PR correlation, introduced in \cite{popescu}, has got large attention in recent years to understand quantum non-locality. This correlation is a bipartite non-signaling correlation which achieves the maximum algebraic value of Bell-CHSH expression. If $A$ and $B$ are the binary input of two distance parties with binary outputs $a$ and $b$ respectively, then PR correlation is describe as
\begin{eqnarray}
P(ab|AB)&=&\frac{1}{2},~~if~a\oplus b=AB\nonumber\\
&=&0,~~otherwise\nonumber\\
\end{eqnarray}
here $A$, $B$, $a$, $b$ take values from $\{0,1\}$. In \cite{oppenheim} it has pointed out that $\zeta_{PR_{cond}}=1$ for conditional (collapsed) distributions of a PR box correlation. We can therefore conclude that conditional distributions of PR correlation will not allow any genuine random process.

\emph{Spekkens's toy theory} : Introducing a foundational principle, namely \emph{knowledge balance principle}, Spekkens constructs a toy theory \cite{spekkens1} in defense of the epistemic view of quantum states. A wide variety of phenomena are found to be reproduced
within this toy theory analogous to quantum mechanics. Our derived relation certifies a genuine random process in this toy theory where the minimum amount of intrinsic randomness guaranteed by this process differs from that of quantum mechanics. It can be shown that in the framework of fine-grained formalism the amount of uncertainty present in toy theory amounts to $\zeta_{toy}=\frac{3}{4}$, therefore there exists a genuine random process which certifies at least $0.83$ bits of intrinsic randomness in toy theory.

\section{ Discussion}
Quantum mechanics, till date, is the most successful theory to describe physical world. There exists various different aspects like intrinsic randomness \cite{Benioff}, uncertainty, nonlocality, steering, entanglement \emph{etc.} that make fundamental distinction between quantum mechanics and classical mechanics. But all these aspects and possible relations among them are not yet well understood from very foundational perspective. Various interesting results have been proved recently concerning these issues, particularly about randomness, nonlocality and uncertainty \cite{oppenheim,pironio,manik,colbeck,gallego,acin}. Pironio \emph{et. al.} have showed that Bell's theorem can certify random numbers \cite{pironio}. In \cite{colbeck,gallego}, it has been proved that intrinsic randomness can be amplified. Acin \emph{et. al.} discussed about possible connection between randomness and nonlocality \cite{acin}. The question of intrinsic randomness attracts so much interest as it has practical importance in various areas like cryptography, gambling, numerical and biological simulations using monte-carlo method \emph{etc}. But mathematical difficulties of characterizing random numbers \cite{knuth} force us to look for physical process where generation of random number can be relied on unpredictability of that physical event \cite{stefanov,dynes,atsushi}. So the question of existence of genuine random process in a particular theory demands practical importance along with foundational interest. In this present article, we have pointed out that though Born's rule is considered as one of the milestone postulate which has introduced intrinsic randomness in fundamental physics, it cannot certify quantum measurement process as \emph{genuine source of intrinsic randomness} in quantum mechanics. On the other hand if uncertainty principle is taken as granted, then \emph{genuine source of intrinsic randomness} can be certified not only in quantum mechanics but in all probabilistic theory. We also derive a quantitative connection between amount of uncertainty and minimum amount of intrinsic randomness generated from a genuine random source in any GNST. In \cite{oppenheim} it has been proved that in any probabilistic theory the amount of nonlocality is determined by the strength of uncertainty accompanied with the strength of steering. In view of this result we can say that the minimum amount of genuine randomness certified in a single party system of a GNST, alone, cannot quantify the amount of nonlocality in a bipartite system of the concerned GNST.

Our finding establishes a fundamental quantitative link between two different aspects, namely  intrinsic randomness and uncertainty, of any general theory and opens few interesting questions. First of all it is worth interesting to find whether preparation state independent intrinsic randomness can be guaranteed and quantified by complementarity principle, another important feature of quantum mechanics. Our intuition go affirmative in this case. It is also interesting to study whether it is possible to quantify preparation state independent intrinsic randomness, considering other forms of uncertainty relation.

\section*{Acknowledgments :}We like to thank G.Kar for many simulating discussion and giving suggestions. TC thanks Council of Scientific and Industrial Research, India for financial support through Senior Research Fellowship (GrantNo. 09/093(0134)/2010). MB like to acknowledge discussion with A. Rai, Md. R. Gazi and S. Das. PP thank Council of Scientific and Industrial Research, India, for financial support.



\end{document}